# Compendium for precise ac measurements of the quantum Hall resistance

F J Ahlers[1], B Jeanneret[2], F Overney[2], J Schurr[1] and B M Wood[3]

[1] Physikalisch-Technische Bundesanstalt (PTB), D-38116 Braunschweig, Germany
[2] Federal Office of Metrology (METAS), CH-3003 Bern-Wabern, Switzerland
[3] Institute for National Measurement Standards, National Research Council (NRC), Ottawa, ON K1A OR6, Canada

**Abstract.** In view of the progress achieved in the field of the ac quantum Hall effect, the Working Group of the Comité Consultatif d'Électricité et Magnétisme (CCEM) on the AC Quantum Hall Effect asked the authors of this paper to write a compendium which integrates their experiences with ac measurements of the quantum Hall resistance. In addition to the important early work performed at the Bureau International des Poids et Mesures and the National Physical Laboratory, UK, further experience has been gained during a collaboration of the authors' institutes NRC, METAS, and PTB, and excellent agreement between the results of different national metrology institutes has been achieved. This compendium summarizes the present state of the authors' knowledge and reviews the experiences, tests and precautions that the authors have employed to achieve accurate measurements of the ac quantum Hall effect. This work shows how the ac quantum Hall effect can be reliably used as a quantum standard of ac resistance having a relative uncertainty of a few parts in $10^8$.

## 1. Introduction

The integer quantum Hall effect (QHE) corresponds to the quantized Hall resistance, $R_H = R_K/i$ with the quantum number $i = 1, 2, 3, ...$ and $R_K$ being the von-Klitzing constant, and to the simultaneous vanishing of the longitudinal resistance. The quantum Hall resistance (QHR) corresponding to the quantum number $i = 2$ and measured with direct current (dc) can be reproduced with an accuracy of 1 part in $10^9$, or better [1]. Following the recommendation of the CIPM (Comité International des Poids et Mesures), the dc QHR has been used since 1990 as the representation of the electrical resistance unit [2,3]. This success has stimulated research into the QHE with alternating current (ac) at frequencies in the kHz range with the aim to de-velop a quantum standard of impedance and to establish a realization of the capacitance unit [4].

After the first precision ac measurement of the QHR had demonstrated an accuracy of a few parts in $10^6$ [4], considerable progress has been made on the subject of accurate ac QHR measurements [5- 15]. Today, measurements with carefully set up ac bridges achieve an uncertainty of a few parts in $10^8$, and agreement at this level has been achieved between different NMIs. In this compendium we summarize this work and the procedures which allow reproducing an ac-QHR measurement with such uncertainty. In this way the QHE can directly be used as a quantum standard of impedance, without resorting to calculable resistor artifacts. This quantum standard can be used to derive the capacitance unit from the von-Klitzing constant, without need of a calculable capacitor. Furthermore, the quantum Hall effect can be exploited at dc, as well as at ac frequencies using one device in one appropriately designed cryomagnetic set-up.

Just as in the dc case, a key quantity for a reproducible, low-uncertainty ac measurement of the quantized Hall resistance is the longitudinal resistance and the understanding of how to control it. The main part of this compendium is devoted to describing the appropriate ac measurement instrumentation (section 3), and the procedures that were used to assure a vanishing longitudinal resistance (section 4).

To keep this compendium within a reasonable length, some subjects that are well described elsewhere are not detailed here. These include: theoretical aspects of the QHE with dc and ac, manufacturing techniques and characteristics of GaAs/AlGaAs heterostructures, design and operation of coaxial ac bridges, and the analysis of the multiple-series connection scheme of QHR devices. Details of coaxial ac bridges in combination with QHE devices will be published separately [16].

## 2. The difference between dc and ac QHE

The key difference between the ac case and the dc case are capacitances that exist between the QHE device and its environment and within the QHE device itself. They cause frequency- and current-dependent effects which lead to apparent differences between the ac QHE and the dc QHE, namely curved Hall plateaus with frequency- and current- dependent deviations from the quantized dc value. The capacitive currents, although small in size, flow through the comparatively large impedance of the QHR and give rise to significant additional voltages in series with the expected Hall voltage across the QHR device. The capacitances would be of minor importance if they would only contribute to the imaginary part of the QHR. However, significant loss factors are associated with these capacitances so that the real part of the ac QHR does exhibit a frequency- and current-dependence.

In fact, the frequency- and current-dependent superposed effects are now well understood and can be largely controlled. A development that is key to this progress is the ability to measure the ac longitudinal resistance, $R_{xx}$, on both sides of the QHE device [10,11]. The ac $R_{xx}$ senses the capacitive effects, and its measurement has considerably improved our understanding of these effects.

Another reason for this progress was an improved understanding of the effect of gates [6-8]. In general, a gate, and other conductors close to the QHE device, act as parasitic capacitive electrodes and increase the frequency- and current-dependent effects [8]. An approximate solution to this intrinsic problem is to avoid all metals close to the QHE device and then to correct for the internal and the residual external effects [5,12]. This is the core of the first method de-scribed below, which extrapolates a series of measurements with finite ac $R_{xx}$ values to the limit $R_{xx} = 0$ and to thus obtain the limit value for $R_{xy}$ with low uncertainty.

In contrast, the second method uses a deliberately made gate structure close to the QHE device. Applying an adjustable ac voltage (of the same frequency) to the gate, the capacitive effects are nulled [6,13], which is verified by monitoring the vanishing of the current dependence of the Hall voltage $R_{xy}$ in the central plateau region. Careful comparisons have shown that both approaches practically give the same ac resistance value and that this value agrees with the dc resistance value, both within an uncertainty of a few parts in $10^8$.

## 3. QHE devices and experimental set-up

### 3.1. QHE devices, cryo-magnetic setup

QHR measurements with ac can, and indeed should be, performed with the same type of devices as used at dc. Selection criteria are discussed in [3], the *'Revised technical guidelines for reliable dc measurements of the quantized Hall resistance'*. Only a device complying with all dc requirements listed in [3] should be used at ac. In all work reviewed in this compendium GaAs/AlGaAs heterostructures were used which either had been manufactured at the Laboratoire Électronique Philips (LEP) by means of MOCVD[*] [17] or at PTB by means of MBE[†] [18,19]. Devices from others sources, if they fulfill the criteria in [3], should of course lead to the same results as reviewed below. In fact, one of the motivations for this compendium was to provide other research groups with a detailed account of the procedures enabling such results.

Similar to the device selection criteria, the cryo-magnetic setup requirements are comprehensively described in [3]. In addition we recommend a cryostat insert which can be evacuated to a sufficiently low residual pressure before the QHR device is cooled down. This minimizes the formation of adsorbates on the device and the associated ac losses in such a layer which can affect a sensitive ac-QHR measurement..

[*] Metal Organic Chemical Vapour Deposition
[†] Molecular Beam Epitaxy

The QHE devices should be operated cautiously and the ac current should be switched on and off with a smooth ramp. Otherwise a large current pulse could flow through the device and this may cause irregular structures in the plateaus (probably due to frozen-in charge carriers). Such structures usually increase linearly with frequency so that ac QHE measurements are more sensitive than dc QHE measurements. However, in the same way as for dc QHE devices, restoration of the original state can always be achieved by cycling the device through room-temperature.

With respect to all other precautions, like the cooling procedure of the devices, the guidelines given in [3] should be followed. In the following we address only those issues which pertain to ac-measurements.

### 3.2. Impedance bridges, device connection scheme

A ratio bridge allows calibrations of an ac resistor in terms of the QHE. A simple scheme of a 1:1 ratio bridge is shown in figure 1b). A four terminal-pair connection scheme [23] as usually applied to conventional impedance standards would give rise to systematic and significant frequency-dependent effects and can not be used straightforwardly in the ac QHR case. The multiple-series connection scheme [10,20,21] shown in figure 1a) elegantly avoids these problems and is used by most NMIs, and especially in all work reviewed in this paper. According to this scheme, all potential contacts of the QHE device which are at the same high or low potential are connected to one junction at high or low potential. These junctions are usually outside of the cryostat at room temperature, and the composite impedance they form can be con-nected to a two or four terminal-pair coaxial bridge. The multiple-series connection can also be used to realize a quadrature bridge [23] as shown in figure 2 with two QHR devices,. Such a bridge design can be used to calibrate capacitors directly in terms of the QHR.

For the bridges shown in figures 1b and 2 the balance conditions for the signal are given by:

$$\frac{R_1}{R_2} = 1 + \alpha + \mathbf{j}\beta\omega C R_2 \tag{1}$$

$$\omega^2 R_1 R_2 C_1 C_2 = 1 + \alpha\, C_4 / C_2 + \mathbf{j}\beta\omega C_3 R_1 \tag{2}$$

where the indexed quantities $R$ and $C$ refer to the resistors and capacitors in the figures and $\omega$ is the bridge operation angular frequency. $\alpha$ and $\beta$ (generally of different value in equation 1 and 2) are adjusted in order to balance the bridge and are sometimes referred to as 'balance parameters'.

### 3.3. Device mounting

The crucial role of the capacitive device environment requires special attention when mounting the device. The choice we recommend is the standardized mounting system used in EU-ROMET project 540 [8]. That system employs a metal case and coaxial leads. The QHE device is glued on a printed circuit board (PCB) inside the case. Pins on the PCB and corresponding sockets in the case allow a simple exchange of QHE devices, and line conductors from the pins close to the QHE device allow short bond wire connections to the contacts of the QHE device.

Securing the QHE device by glue avoids ac driven vibrations, and for the same reason the bond wires should be kept short [20]. Only a small amount of glue should be used, even if the loss factor of the glue at low temperature is very small. The current contacts from the device to the PCB can be provided with two parallel bond wires of different lengths to prevent coinciding resonance frequencies.

The metal case of the device holder is part of the outer-conductor network of the coaxial bridge and isolated from the cryostat. This prevents capacitive current leakage to the cryostat ground which would render the bridge non-coaxial and the impedance of the QHR ill-defined.

The outer conductors of all coaxial leads to the QHE device must be connected to a central star-point close to the QHE device; otherwise the hindering of coaxial return currents would lead to a large quadratic frequency-dependence. From an electro-technical point of view, the star point, usually a small metal pad, should be close to the QHE device. However, due to capacitive coupling to the device, a metal pad located too close to an ungated QHE device acts as a gate and increases the frequency- and current-dependence of the ac QHR [8]. This can be avoided by arranging the star-point several millimeters from the QHR device. For the same reason, any other metal too close to an ungated QHE device should be avoided.

For the method based on a split gate [6], this split-gate is usually part of the PCB and that PCB, like the one in the commonly employed EUROMET design  [22] , needs to be modified accordingly. Lithographically defined gates on top of the heterostructure and to the side to the Hall bar (so called in-plane gates) have also been used. However, they yield no real advantage and require specially fabricated devices whereas existing standard dc devices can always be mounted onto a PCB with a split gate.

*3.4. Coaxial leads*

The coaxial leads connected to the ac QHR device (and also the strip conductors on the PCB) must exhibit a sufficiently high leakage resistance between the inner and outer conductors ($> 10^8 \cdot R_H = 10^{12}\,\Omega$ for the defining high-potential lead). For gated devices, the leakage resistance between the gate and the QHR device must also be sufficiently high. Both can be tested using a dc teraohm meter.

In general, the outer-conductor network of the multiple-series leads has a four terminal-pair impedance *in parallel* to the QHR. The current equalizers transform this impedance into the inner conductor *in series* with the QHR. In a dc QHR measurement, the current flows only in the inner conductor and the parallel resistance of the outer-conductor network does not affect the dc measurement. Therefore, an ac QHR measurement differs from the dc QHR measurement by this parallel four terminal-pair resistance. Usually this parallel resistance can be made sufficiently small but to safely exclude a systematic error it should be measured using a 4-wire dc microohm meter. If it is not sufficiently small ($< 10^{-8} \cdot R_H \approx 100 \ \mu\Omega$) a correction can then be applied.

The outer conductors of the coaxial leads and the case of the QHR device must provide sufficient screening to avoid any spurious crosstalk between the inner conductors of different leads or between each inner conductor and any other device, for example, the cryo-magnetic system. The screening can be tested with a high-resolution ($\approx 0.1$ fF) capacitance bridge.

The usual cable corrections accounting for the impedance and admittance of the defining leads [23] in two- or four-terminal-pair impedance measurements also need to be applied to the multiple-series connection of the QHR (see for example, [10]). The coaxial leads are normally composed of sections with different impedance and admittance per meter (for example, a room-temperature cable, a cryogenic cable, and the strip line on the printed circuit board) and may require a more complex cable correction.

In order to determine the cable parameters with an LCR meter, two dummy printed circuit boards without a QHR device should be used: one printed circuit board with open terminals for the measurement of the lead admittance, and a second printed circuit board with short-circuited terminals for the measurement of the lead impedance. The cable parameters should be measured in the cryo-magnetic system under the same conditions as for the ac QHR measurement.

### 3.5. AC contact resistance

Contact to the 2DEG is achieved by bonding wires to lithographic AuGeNi contacts or by soldering wires to tin-ball contacts. Non-ohmic or high value contacts can cause a systematic shift of the quantum Hall resistance and/or distorted plateau shapes. It is therefore necessary to characterize the contacts. A suitable three terminal-pair measurement of the ac contact resistances is described in [10] and is simple, quick and has a resolution well below 0.1 $\Omega$.

The ac current used for the measurement of the current-contact resistance should be similar to that of the ac QHR measurements. The measurement current applied to the potential contacts should be considerably smaller because the potential contacts carry only a very small current during an ac QHR measurement, and in the case of lithographically manufactured AuGeNi contacts the potential contacts usually have smaller dimensions and tolerate less current.

The contact resistances can be measured at different frequencies. If resonances occur, this may indicate mechanical vibrations of excessively long bond wires. Usually, the ac contact impedances are resistive and show no frequency dependence so that the ac and dc contact resistances are virtually the same.

The resistance of good ohmic contacts is usually less than 1 $\Omega$. Large contact resistances ($> 50\ \Omega$) can cause two effects. First, the small currents flowing in the potential leads as a consequence of the multiple-series connection produce a small, calculable voltage drop at the potential contacts which must be taken into account [10]. The second possible effect of poor contacts is to produce a deviation of the quantum Hall resistance from its quantized value, $R_{\mathrm{K}}/i$, and this is often accompanied by excess noise and irregular structures in the plateau. In one example of a poor contact resistance we found that even at a resistance value of 50 $\Omega$ the ac quantum Hall resistance deviated from the quantized value by less than 1 part in $10^8$. Though this was only a single result it does suggest that the contact resistance requirements in the ac case are very similar to those of the dc case [2,3].

### 3.6. Coaxial bridges

The basic techniques of coaxial ac bridges are comprehensively described in [23]. Different bridges have yielded the same, correct results - provided that the bridges satisfy the basic principles. A summary of current experiences with coaxial ratio bridges, particularly in combination with QHE devices, will be published in [16]. Coaxial bridges for the measurement of the ac longitudinal resistance on both sides of the QHE device and for the measurement of the ac contact resistance have been developed and are described in [10,16].

Coaxial bridges allow measurements with very low uncertainties ($< 1 \cdot 10^{-8} \cdot R_\mathrm{H}$) but they are susceptible to systematic errors. Therefore, the bridges must be carefully designed to meet the defining conditions strictly, and must be extensively tested. This statement applies not only to quadrature and ratio bridges but also to 10:1 ratio calibration and $R_\mathrm{xx}$ bridges. A detailed and complete uncertainty budget [24] for each bridge must be determined. To avoid overly optimistic estimates, we recommend the determination of individual contributions experimentally.

### 3.7. Current equalizers

Current equalization is an important requirement of the coaxial ac technique [23]: the current flowing in the outer conductor of each coaxial cable must be equal and opposite to the current in the inner conductor. Equalized currents make the bridge network immune to external magnetic and electric fields so that mains and radiofrequency interferences are eliminated. Fur-thermore, the current in the network does not generate external magnetic and electric fields which could significantly influence the bridge balance condition.

To achieve equal return currents in the whole bridge network, each independent mesh of the bridge network must be provided with exactly one current equalizer [23]. Otherwise, the bridge network is susceptible to interference signals, ac currents flow along unexpected paths, and fractions of the large impedances of the current equalizers are transformed into the network of the inner conductors and cause systematic errors [25]. Therefore, any coaxial bridge must be properly equalized and, since each individual current equalizer may still have a significant effect on the main balance, the individual effects must be determined experimentally. If their sum is not small enough, a correction must be applied [23].

The ac QHR coaxial leads in the cryostat are often long and thin and thus the outer conductors have comparatively large impedances which reduce the efficiency of the current equalizers and increase their effect on the main balance. The equalizer efficiency can be improved with active equalizers employing a feedback system which injects a current into the outer conductor proportional to the measured current inequality [26].

Not all equalizers in a bridge network affect the main balance in the same way. The ones with a significant effect are usually replaced by active equalizers. An indiscriminate use of active current equalizers will, however, not help if the basic principles of coaxial bridges are not carefully met. Even with active equalizers it is still necessary to measure the net currents throughout the whole bridge network and to evaluate the total effect of all current equalizers on the main balance.

*3.8. DC measurements of the ac QHR*

An ac QHR device mounted in a cryo-magnetic system with coaxial leads can also be used for dc QHR measurements. The four dc measurement leads can be directly connected either to the corresponding coaxial sockets at the top of the cryostat insert or to the two multiple-series junctions outside of the cryo-magnetic system. In this way, the dc Guidelines [3] can be evaluated. In fact, some measurements apply to ac and dc in the same way, for example the measurements of the contact resistances and of the leakage resistance of the output leads. Further, ac $R_{xx}$ measurements also yield a value for the dc $R_{xx}$ (or at least an upper limit). Therefore, it is possible and reasonable to use one QHR device for both ac and dc measurements.

## 4. The QHR as a primary impedance standard

*4.1. Frequency dependence of longitudinal and Hall resistance*

An ac measurement of the QHE which does not account for the capacitive effects and the losses they cause results in deviations $\Delta R_H := R_H(f,I)$ - $R_K/i$ of $R_H$ from the quantized dc value. Further the $R_{xx}(B)$ and $R_H(B)$ plateaus (where $B$ is the magnetic induction) are not as flat as measured with dc, and they are usually curved or show some other kind of structure. The effects increase approximately linearly with frequency and current, in the frequency range up to at least 10 kHz and for currents up to a few tens of μA.

Extensive experimental investigations of these effects have been carried out. Most were done on the plateau at filling factor $i = 2$, at a temperature of 0.3 K, and with devices showing a perfect dc quantization (i.e. $\rho_{xx}^{dc} = 0$). A typical result of a measurement of the Hall resistance $R_H$ is shown in figure 3. The longitudinal resistivity $\rho_{xx}$ and $\Delta R_H$ are shown with dependence on current and frequency in figure 4. They were obtained at METAS and at PTB, using the same device. Despite the use of different coaxial measuring bridges, the results show excellent agreement, and they can be approximated by the following relations:

$$\frac{\rho_{xx}}{R_H} \approx f \cdot (k + c \cdot I) \qquad (3)$$

$$\frac{\Delta R_H}{R_H} \approx f \cdot (k' + c' \cdot I) \qquad (4)$$

These frequency- and current-dependences are characteristic for the ac QHE. The four slope parameters $c$, $c'$, $k$ and $k'$ depend on the exact capacitive environment of the device and are in the following approximate proportion $k/c \approx k'/c'$.

The linear frequency- and current-dependence applies not only to the values at the plateau center but also to the plateau shape. All details observed in the plateau shape of $\Delta R_H$ are replicated in the shape of $\rho_{xx}$. The similarity of $\Delta R_H$ and $\rho_{xx}$ indicates that the same physical process is involved, and that this can be used to eliminate this effect. In the following, two methods will be discussed which accomplish such an elimination. The associated models which will be presented in section 5 not only describe the main findings of each method and complement each other but also make it clear that both methods can be interpreted in the same way and are thus equivalent (section 5.3).

*4.2. Measurement with ungated devices*

In experimental set-ups where no conductor at shield potential is close to the device, an extrapolation method can be used to obtain the correctly quantized Hall resistance in an ac measurement. The basis of the method is an experimentally found linear correlation between $\Delta R_H$ and $\rho_{xx}$ illustrated in figure 5 and expressed by

$$\Delta R_{\mathrm{H}} \approx s' \cdot \rho_{\mathrm{xx}}(f, I) = (s - g) \cdot \rho_{\mathrm{xx}}(f, I) \tag{5}$$

The proportionality factor $s'$ is written as a difference of two contributions, $s$ and $g$, although this distinction is not vital for the applicability of the extrapolation method. When $\Delta R_{\mathrm{H}}$ is measured in different geometrical directions by using different potential contacts, a different geometrical parameter $g$ leads to a different superposition of $\rho_{\mathrm{xx}}$ on $\Delta R_{\mathrm{H}}$. This explains why some diagonal measurements of the Hall resistance, where incidentally $g \approx s$, yield a small residual current- and frequency-dependence [5] while others with $g > s$ yield a negative current- and frequency-dependence [12]. The longitudinal resistivity $\rho_{\mathrm{xx}}$ was observed to always show a positive current- and frequency-dependence ($c > 0$). The nongeometrical factors $s$ of different QHE devices have typically values of about 2. While (5) is based on experimental observations, its validity is substantiated by a phenomenological model using an electromagnetic argument which is described in section 5.1.

Relation (5) is the basis of the extrapolation method because it tells us that the measured ac Hall resistance equals $R_{\mathrm{K}}/i$ to the same degree that $\rho_{\mathrm{xx}} \approx 0$ is fulfilled. This is basically the same condition as in a dc measurement [1-3]. When applying the method, $\rho_{\mathrm{xx}}$ is measured at different currents at a fixed frequency, and $R_{\mathrm{H}}$ is measured at those same currents and frequency. Then one extrapolates $R_{\mathrm{H}}$ to the condition where $\rho_{\mathrm{xx}} \approx 0$ is fulfilled. The method does not need a reference resistor with a calculable frequency dependence. In fact, (5) can be used for the calibration of an unknown impedance standard, $R_{\mathrm{DUT}}$, at kHz frequencies.

The detailed procedure as described in [12] is as follows. For impedance ratio bridges, the ratio of the compared impedances is derived from (1) as

$$\frac{R_{\mathrm{H}}(f, I)}{R_{\mathrm{DUT}}(f)} - 1 = \alpha(f, I) \tag{6}$$

where $\alpha(f,I)$ is the in-phase parameter needed to balance the bridge at frequency $f$ and current $I$. For a given frequency $f_{\mathrm{n}}$, the bridge is balanced differently for different currents $I_{\mathrm{i}}$ (i = 0 to $N$) leading to a series of balance parameters $\alpha(f_{\mathrm{n}}, I_{\mathrm{i}})$. The unknown impedance $R_{\mathrm{DUT}}$ is assumed to be current independent for the current range used[‡].

The longitudinal resistivity $\rho_{\mathrm{xx}}(f_{\mathrm{n}}, I_{\mathrm{i}})$ is measured at the same frequencies and currents, for simplicity along the low-potential edge of the device. The locus of the points $\{\rho_{\mathrm{xx}}(f_{\mathrm{n}}, I_{\mathrm{i}}), \alpha(f_{\mathrm{n}}, I_{\mathrm{i}})\}$ in the $\{\rho_{\mathrm{xx}}, \alpha\}$ plane is a straight line (see figure 6) represented by

$$\frac{R_{\mathrm{K}}/2 + s' \cdot \rho_{\mathrm{xx}}(f_{\mathrm{n}}, I_{\mathrm{i}})}{R_{\mathrm{DUT}}(f_{\mathrm{n}})} - 1 = \alpha(f_{\mathrm{n}}, I_{\mathrm{i}}) \tag{7}$$

where the Hall resistance $R_{\mathrm{H}}$ takes the value $R_{\mathrm{K}}/2$ on the usually employed QHE plateau at filling factor 2. A least-squares fit of a line to the $\{\rho_{\mathrm{xx}}, \alpha\}$ points gives the slope $s'$, as well as $\alpha_0(f_{\mathrm{n}})$ which is the value of $\alpha$ extrapolated to zero longitudinal resistivity (zero dissipation).

$$\alpha_0(f_{\mathrm{n}}) = \lim_{\rho_{\mathrm{xx}} \to 0} \frac{R_{\mathrm{K}}/2 + s' \cdot \rho_{\mathrm{xx}}(f_{\mathrm{n}}, I_{\mathrm{i}})}{R_{\mathrm{DUT}}(f_{\mathrm{n}})} - 1 \tag{8}$$

In other words, $\alpha_0(f_{\mathrm{n}})$ is the balance one would obtain if there was no dissipation in the quantum Hall device. It is worth mentioning that the limit of zero dissipation is not equivalent to the limit of zero current. Indeed, $\lim_{I \to 0} \rho_{\mathrm{xx}}(f_{\mathrm{n}}, I_{\mathrm{i}})$ is not necessarily zero! Finally, (8) leads to:

$$R_{\mathrm{DUT}}(f_{\mathrm{n}}) = \frac{R_{\mathrm{K}}}{2}\left(1 - \frac{\alpha_0(f_{\mathrm{n}})}{1 + \alpha_0(f_{\mathrm{n}})}\right) \approx \frac{R_{\mathrm{K}}}{2}\left(1 - \alpha_0(f_{\mathrm{n}})\right) \tag{9}$$

---

[‡] If this is not the case, it is necessary to determine the current dependence separately.

This procedure has successfully been applied to the calibration of the frequency dependence of a quadrifilar calculable resistor [27], and within the expanded uncertainty ($k = 2$) of the measurement (about 0.05 $\mu\Omega/\Omega$ at 1 kHz) [12] no difference between the measured and calculated frequency dependence was found as can be shown in Fig.7.

### 4.3. Gating method

The second method to remove the frequency- and current-dependence of $R_\mathrm{H}$ uses QHR devices with two gates, usually in the form of a split back-gate on the PCB which carries the QHR chip [6,13]. The two half gates are provided with two adjustable ac voltages $U_\mathrm{GHi}$ and $U_\mathrm{GLo}$ of the same frequency as the Hall voltage $U$. The frequency- and current-dependence of $\rho_\mathrm{xx}$ and $R_\mathrm{H}$ can be empirically written as

$$\frac{\rho_\mathrm{xxHi}}{R_\mathrm{H}} \approx f \cdot \left(k_\mathrm{Hi} + c_\mathrm{Hi} \cdot I\right) \cdot \lambda_\mathrm{Hi} \tag{10}$$

$$\frac{\rho_\mathrm{xxLo}}{R_\mathrm{H}} \approx f \cdot \left(k_\mathrm{Lo} + c_\mathrm{Lo} \cdot I\right) \cdot \lambda_\mathrm{Lo} \tag{11}$$

$$\frac{\Delta R_\mathrm{H}}{R_\mathrm{H}} \approx f \cdot \left(k'_\mathrm{Hi} + c'_\mathrm{Hi} \cdot I\right) \cdot \lambda_\mathrm{Hi} - f \cdot \left(k'_\mathrm{Lo} + c'_\mathrm{Lo} \cdot I\right) \cdot \lambda_\mathrm{Lo} \tag{12}$$

The frequency- and current-dependent effects along the high- and low-potential edges are now distinguished, and it is noted that the contributions to $\Delta R_\mathrm{H}$ from the high- and low-potential edges have opposite sign.

If the gate arrangement is symmetrical and the loss mechanisms at both edges of the device are the same, the frequency- and current-coefficients at both sides of the device have approximately the same ratio:

$$k'_\mathrm{Hi}\Big/c'_\mathrm{Hi} \approx k'_\mathrm{Lo}\Big/c'_\mathrm{Lo} \tag{13}$$

The presence of gates driven at an ac voltage $U_G$ (or of any conductor at shield potential close to the device) modifies the frequency- and current-coefficients in a way which can be described by factors $\lambda$ . These factors describe how much the capacitive loading of the QHR device is reduced by biasing the capacitances of the QHR device with the external electric gate field [7]. This will be further discussed in section 5.2. Here it is only relevant that the effect of gates can be described by a multiplication (and not by an additional sum term), which is evident from the results in figure (8a) . For an ungated device, $\lambda$ equals one.

As described by (10) to (12), gates allow control of the frequency- and current-dependent effects. The adjustment procedure is as follows: The magnetic field is set to a value not too far from the center of the $R_H$ plateau. The two gate voltages are changed until the current coefficient of $R_H$, $dR_H/dI$, becomes zero . Such a reduction of the current coefficient to zero can be achieved because the contributions to $R_H$ from the two sides of the device have opposite sign and cancel each other mutually. In this condition the frequency-dependence also becomes zero [6,12] by virtue of (13).

A mutual compensation of the effects at both sides of the QHR device can be achieved for several combinations of the two gate voltages, which lie on a straight line in the two-dimensional gate-parameter space. The mutually compensating effects on both sides of the device often have a slightly different dependence on the magnetic field so that the remaining Hall plateau is not perfectly flat. This does not limit the gate method and shows that a flat plateau is neither necessary nor sufficient for ac measurements.

At one special setting of the gate-voltages with a zero current coefficient, both $\rho_{xx}$ and $\Delta R_H$ simultaneously become zero and perfectly flat over a broad range of the magnetic field, independent of frequency and current [6,12]. For this special setting, both $\lambda_{Hi}$ and $\lambda_{Lo}$ in (10) to

(12) are zero. Now the effects at both sides of the device are individually zero (and do not require a mutual compensation). Figure 8 illustrates such a nearly perfect adjustment of the $R_{xx}$ and $R_H$ plateaus. Again we like to emphasize that the flatness of the $R_H$ plateau as it occurs at this special gate setting is not a necessary requirement for a correct $R_H$ measurement. Since this special gate setting requires ac bridges measuring $\rho_{xx}$ on both sides of the device [10,11], we recommend it as the simplest procedure to null the current coefficient of $R_H$ since this is sufficient to ensure a zero frequency coefficient of $R_H$, and thus a vanishing deviation of $R_H$ from the quantized dc value.

In conclusion, gated QHR devices allow the realization of a quantum standard of impedance, independent of a classical resistance artifact with calculable frequency dependence.

### 4.4. Magnetic field reversal

Reversing the polarity of the magnetic field and/or interchanging the high and low potential connections of the QHR device are necessary tests in the dc case according to [2,3]. For a reversal of the magnetic field in the ac case the potential leads have to be re-configured because of the multiple-series connection scheme which implies that the ac QHR is measured at different potential contacts and in a different geometrical direction. For this reason, the different magnetic field configurations will in general require slightly different gate settings or yield slightly different extrapolation parameters, but both methods provide consistent values of the quantum Hall impedance.

## 5. Model for ac losses and equivalent circuit

While we reviewed in the previous section two methods for eliminating the frequency- and current-dependence of the ac QHE, we summarize in this section an empirical modeling which underlies these methods and which represents our understanding of the ac QHE. We describe this modeling in two formulations which complement each other. In one (section 5.1), we use a description which emphasizes the physical loss mechanisms in the devices, whereas in the other formulation (section 5.2) we use an electro-technical equivalent circuit which emphasizes the capacitive effects.

### 5.1. Polarization model

At dc, the sheet current density $\vec{J}$ and the electric field $\vec{E}$ are related by the conductivity tensor $\boldsymbol{\sigma}$. At finite frequencies, one must take the displacement field into account:

$$\vec{J}(t) = \sigma \cdot \vec{E}(t) + d\frac{\partial \vec{D}(t)}{\partial t} \qquad (14)$$

The last term in (14) represents the sheet density of displacement current $\vec{J}_D$. The coefficient $d$ is the thickness of the layer in which this current flows. Part of $\vec{J}_D$ can flow in the 2DEG itself and part of it can certainly flow in the various semiconducting layers adjacent to the 2DEG. An important point is that the main result of this model does not depend on the particular distribution of $\vec{J}_D$.

The displacement field $\vec{D}$ is related to the polarization $\vec{P}$ by:

$$\vec{D}(t) = \varepsilon_0 \cdot \vec{E}(t) + \vec{P}(t) \qquad (15)$$

and, assuming a linear response, the polarization is related to the electric field by:

$$\vec{P}(t) = \varepsilon_0 \chi \cdot \vec{E}(t) \qquad (16)$$

where the dielectric susceptibility, $\chi$, is a complex tensor in the general case (see [28] for more details). Here the polarization is assumed to stay in the same plane as the electric field (2D model) at an angle $\varphi$. This assumption is reasonable in case there are no interactions between the sample and its surroundings (eg. back gates, screening electrodes...). In addition the effective thickness $d$ is much smaller than the sample's width and length.

Solving the above equations gives a result that shows a linear frequency dependence for both the longitudinal and the Hall resistance. The final result for the deviation from perfect quantization is:

$$\Delta R_H = \rho_{xy}^{ac} - R_H = -\tan(\varphi)\rho_{xx}^{ac} = s \cdot \rho_{xx}^{ac} \tag{17}$$

The deviation of the Hall resistance from the perfect quantization is proportional to the dissipation in the system as observed experimentally (see figure 5). This dissipation is caused by the displacement current flowing in the system; a situation completely different from the dc case where thermal activation and variable range hopping were responsible for this non-ideal behavior [1]. The parameter $s$ is positive and in the range between 1.0 to 2.5 for the different samples measured. This parameter is more than one order of magnitude larger in ac measurements than in the dc case, in addition to having the opposite sign.

The most important result of this model is to explain the proportionality factor between the deviation from the perfect quantization of $R_H$ and the dissipation $\rho_{xx}^{ac}$. Using (17), the parameter $s$ can be converted to the angle $\varphi$ between the electrical field $\vec{E}$ and the polarization field $\vec{P}$. The values of $\varphi$ range from -45° to -68° (corresponding to a parameter $s$ in the range 1.0 to 2.5 [28]). A weak variation of the parameter $s$ observed after different room-temperature cycles can be explained by cool-down dependent effects that modify the microscopic charge distribution and so influence the electric susceptibility $\chi$. A dependence of $s$ on the magnetic field direction was also observed.

## 5.2. Capacitive model

The capacitive model is based on the analysis of the equivalent circuit of a QHE device [10]. When a capacitive current flows through the QHE device, the equivalent circuit shows that a corresponding Hall voltage is generated between the particular contacts, and this voltage contributes to the measured longitudinal and Hall resistance. The relevant capacitances are the internal capacitances due to the current distribution inside the device, $C_{int}$, and the external capacitances between the edges of the device and their metallic environment, $C_{ext}$. The microscopic details of the internal capacitances are not fully understood but their effect on the ac longitudinal impedance and the quantum Hall impedance can nevertheless be calculated and measured. Analyzing the equivalent circuit yields for the longitudinal impedance at the high- and the low-potential side of a QHE device, $Z_{xxHi}$ and $Z_{xxLo}$ [10]:

$$Z_{xxHi} = R_{xx0} + (C_{int\,Hi}[j + \tan \delta_{int\,Hi}] - \alpha C_{ext\,Hi}[j + \tan \delta_{ext\,Hi}]) \omega R_H^2 \lambda_{Hi} \quad (18)$$

$$Z_{xxLo} = R_{xx0} + (C_{int\,Lo}[j + \tan \delta_{int\,Lo}] - \beta C_{ext\,Lo}[j + \tan \delta_{ext\,Lo}]) \omega R_H^2 \lambda_{Lo} \quad (19)$$

$$\text{with} \quad \alpha = (U_{GHi} - U)/U \quad \text{and} \quad \beta = -U_{GLo}/U \quad (20)$$

$R_{xx0}$ is the non-capacitive part of the longitudinal resistance which at the plateau edges is non-zero (the thick curve in figure 8a). It would be measured in a dc experiment as well, and is practically frequency independent. This contribution is assumed to be zero at the center of the plateau.

The capacitances and associated loss factors can be measured by means of a capacitance bridge [6,15]. The results show that they both depend on the filling factor and have a curved shape, in agreement with the results shown in figure (7a). Further, the capacitances are found to be independent of frequency and current whereas the loss factors are independent of fre-quency but approximately proportional to current. With these dependences, equations (18) and (19) can be written in the empirical form of (10) and (11), and describe the observed frequency- and current-dependence of $R_{xx}$.

Again using the equivalent circuit of the QHE device [10,12], the ac quantum Hall impedance, $Z_{xy}$, can be calculated and, using the same notation as above, is found to differ from the quantized dc value by:

$$\Delta Z_{xy} = (s_0 - g)\frac{w}{l}R_{xx0} +$$
$$+ (\widetilde{C}_{intHi}[j + \tan\widetilde{\delta}_{intHi}] - \alpha\widetilde{C}_{extHi}[j + \tan\widetilde{\delta}_{extHi}])\omega R_H^2 \lambda_{Hi} -$$
$$- (\widetilde{C}_{intLo}[j + \tan\widetilde{\delta}_{intLo}] + \beta\widetilde{C}_{extLo}[j + \tan\widetilde{\delta}_{extLo}])\omega R_H^2 \lambda_{Lo}$$

$$\quad (21)$$

$\quad w \quad = \quad$ width of the device
$\quad l \quad = \quad$ distance between potential contacts
$\quad g \quad = \quad$ superposition of $R_{xx0}$ if $R_{xy}$ is measured at
$\qquad\qquad\quad$ non-opposing potential contacts

As argued in [6,7], the capacitive effects occur mainly in the edge regions of the QHE device. The edge regions which contribute to $R_{xy}$ differ from the edge regions which are sensed by the measurement of $R_{xx}$. Therefore, the quantities in (21) are provided with tilde symbols, to distinguish them from the quantities in (18) and (19). The ratio of the capacitive terms contributing to $R_{xy}$ and to $R_{xx}$ corresponds to the parameter $s$ mentioned in section 4.2 which has a value of about 2. In contrast, the $s_0$ parameter has a value of -0.1, and this applies to dc and ac measurements as already shown in [13]. Making use of the measured frequency-independence and approximate current-proportionality of the capacitances and associated loss factors, (21) can be written in the empirical form (12) .

In the central plateau region, the 2DEG behaves like a dielectric and, for a QHR device without any gate, the internal and external capacitances load the QHR device. When a QHR device is provided with a split gate and external voltages $U_{GHi}$ and $U_{GLo}$ are applied to the particular half gate, this generates an external electric field which penetrates into the dielectric region of the 2DEG. The electric field biases the internal and external capacitances and reduces the loading of the QHR. This is described by the loading factors $\lambda_{Hi}$ and $\lambda_{Lo}$ [12]. If the external electric field is adjusted to be equal to the mean Hall electric field, the internal and external capacitances do not present a net load to the QHR, corresponding to $\lambda_{Hi} = \lambda_{Lo} = 0$. For a symmetric gate in the same plane as the 2DEG, the loading factors can be calculated and are approximately $\lambda_{Hi} \approx 1 - \alpha w/\gamma$ and $\lambda_{Lo} \approx 1 - \beta w/\gamma$ with $\alpha$ and $\beta$ as defined in equation (20). The quantity $\gamma$ is the distance between each half gate and the adjacent device edge, and $w$ is the width of the device. Note that both loading factors are approximately independent of each other (so that the adjustment procedure converges quickly, in agreement with the experiment) and they can be positive, negative, or zero.

The gate-voltage dependence of $\lambda_{Hi}$ and $\lambda_{Lo}$ explains the approximately straight line in the two-dimensional gate parameter space along which the current coefficient of $R_{xy}$ is zero [6]. When both loading factors are adjusted to zero, all the particular capacitances do not load the QHE device anymore, and the curved plateau shape and the frequency- and current-dependence of $R_{xx}$ and $R_{xy}$ simultaneously vanish, as observed experimentally.

### 5.3. Summary of the models

The models reviewed in sections 5.1 and 5.2 emphasize different aspects but complement each other. The capacitive model emphasizes the internal and external capacitances along both sides of the device but does not explain the physical mechanism behind the loss tangent. The polarization model suggests a physical loss mechanism for the capacitive currents which is able to explain the loss tangent and the linear frequency dependence. Neither model enlightens the microscopic processes in the 2DEG. However, this is not necessary for a successful measure-ment, a situation not much different from the dc QHE case, where to date no model is able to explain all relevant details of a real QHE device on the microscopic level. For instance, the question of where exactly the current flows is still unanswered. This does not prevent us from using the QHE for dc calibrations at an uncertainty level of 1 part in $10^9$, and the ac QHE is in this sense not different from the dc QHE.

Finally, we want to point out the equivalence of the two methods. The extrapolation method is based on the observation that extrapolating $R_H$ to $\rho_{xx} = 0$ yields the correct resistance value $R_K/i$. The value of $R_H$ measured at a fixed frequency and a finite current differs from the extrapolated value by $\Delta R_H = s \cdot \rho_{xx}$. The alternative to extrapolating $\rho_{xx}$ to zero would be to apply a correction by the same amount $s \cdot \rho_{xx}$. This correction can be viewed as a rotation of the direction in which $R_H$ is determined such that the ac losses on both sides of the device mutually compensate each other. On the other hand, a mutual compensation is also the core of the gating method, even though the compensation there is achieved by controlling gates. Further, both methods suppress the contribution of the unwanted ac losses to $R_H$ by making use of $\rho_{xx}$. For all these reasons the two methods are to be considered as equivalent.

At last, it should be mentioned that both approaches presented here still have systematic uncertainties of a few parts in $10^8$ per kHz. This uncertainty, and also the level of agreement achieved between the contributing NMIs, is remarkably good and much better than the situation several years ago. It allows the use of the ac QHE for impedance calibrations at an uncertainty level similar to that of a calculable cross-capacitor without reliance on resistance artifacts with calculable frequency dependences. Further developments and experimental refinements as well as improvements of our understanding are in progress and accordant results were published recently [14,15]. In these papers the present model of the frequency and cur-rent dependences of the quantum Hall effect was addressed and confirmed, and three different loss mechanisms in ac-QHR measurements were identified. A special double-shielded configuration proposed in [14,15] allowed to eliminate the effect of all three loss mechanisms so that the ac QHR became independent of frequency and current, and an excellent relative uncertainty below $1 \times 10^{-8}$ could be achieved which is a significant improvement in the setup for using the ac QHR as a precise quantum standard of impedance. When the new findings from [14,15] will have been confirmed by other NMIs, a future version of this compendium will advise in detail how to achieve the uncertainty of 1 part in $10^8$ or lower for the ac QHR.

## 6. Conclusion

The understanding of the ac QHE has significantly improved in the past years and a status has been reached where the precision, reproducibility, and reliability of the ac QHE at different NMIs are better than, or at least comparable to, that of classical impedance artifacts which provide a calculable transition from ac to dc. Using the approaches reviewed here, the ac QHR can be used to derive a precise quantum standard of capacitance, independent of a calculable cross capacitor. The knowledge acquired, the measuring procedures and the model, as well as a number of precautions, constraints and details have been summarized, so that this compendium may serve as a foundation for other groups who we would encourage to realize such a quantum standard of impedance.

**Acknowledgments.** We are grateful to F Delahaye (formerly with the BIPM) for providing an LEP split-gate device and for valuable discussions. J Schurr thanks B P Kibble (formerly with the NPL) for useful discussions, and for bridge 'tutoring'.

## Figure Captions

**Figure 1**
(a) Triple series connection of a QHE device. The star points of the triple series connection are at room temperature and can be reconfigured. Grey rectangles indicate typical arrangement of ac driven back-gates as used in the method described in section 4.3. (b) Schematic of a resistance ratio bridge.

**Figure 2**
Schematic of a quadrature bridge used to compare unlike impedances. QHR resistances connected in multiple-series according to Figure 1 (a) can be inserted for $R_1$ and $R_2$.

**Figure 3**
Typical frequency dependence of the Hall resistance $R_{xy}$ at filling factor 2. The observed linear dependence is indicated in the plateau center.

**Figure 4**
Results of $\rho_{xx}/R_H$ along the low-potential edge of an LEP- device (a), and for $\Delta R_H$, the deviation of the resistance $R_H$ from its quantized value $R_K/i$ (b), measured at different frequencies and currents both at METAS (open symbols) and at PTB (solid symbols).

**Figure 5**
Correlation plot of $\Delta R_H$, the measured deviation from the Hall plateau center value $R_H$, against $\rho_{xx,}$ both in units of $R_H$.

**Figure 6**
Linear relation between bridge balance parameter $\alpha$ and $\rho_{xx}$ obtained for three different frequencies and two magnetic field directions. The extrapolation method is based on plots of this kind, as decribed in the text.

**Figure 7**
Top graph: Measured (symbols) and calculated (line) frequency dependence of the calculable quadrifilar resistor. Bottom graph: difference between the measured and calculated value. The dashed lines correspond to the total measurement uncertainty (2 ).

**Figure 8**
(a) Traces of $R_{xx}$ (measured along the low-potential edge) for various gate voltages. The optimally tuned $R_{xx} = 0$ case is shown by the thick line. (b) Traces of $\Delta R_H$ versus filling factor $\nu$ for six different devices, each measured after obtaining optimal tuning by the method described in section 4.3. All Hall plateaus are flat to within 2 parts in $10^8$. For clarity all curves are offset by an equal amount.

.

(a)

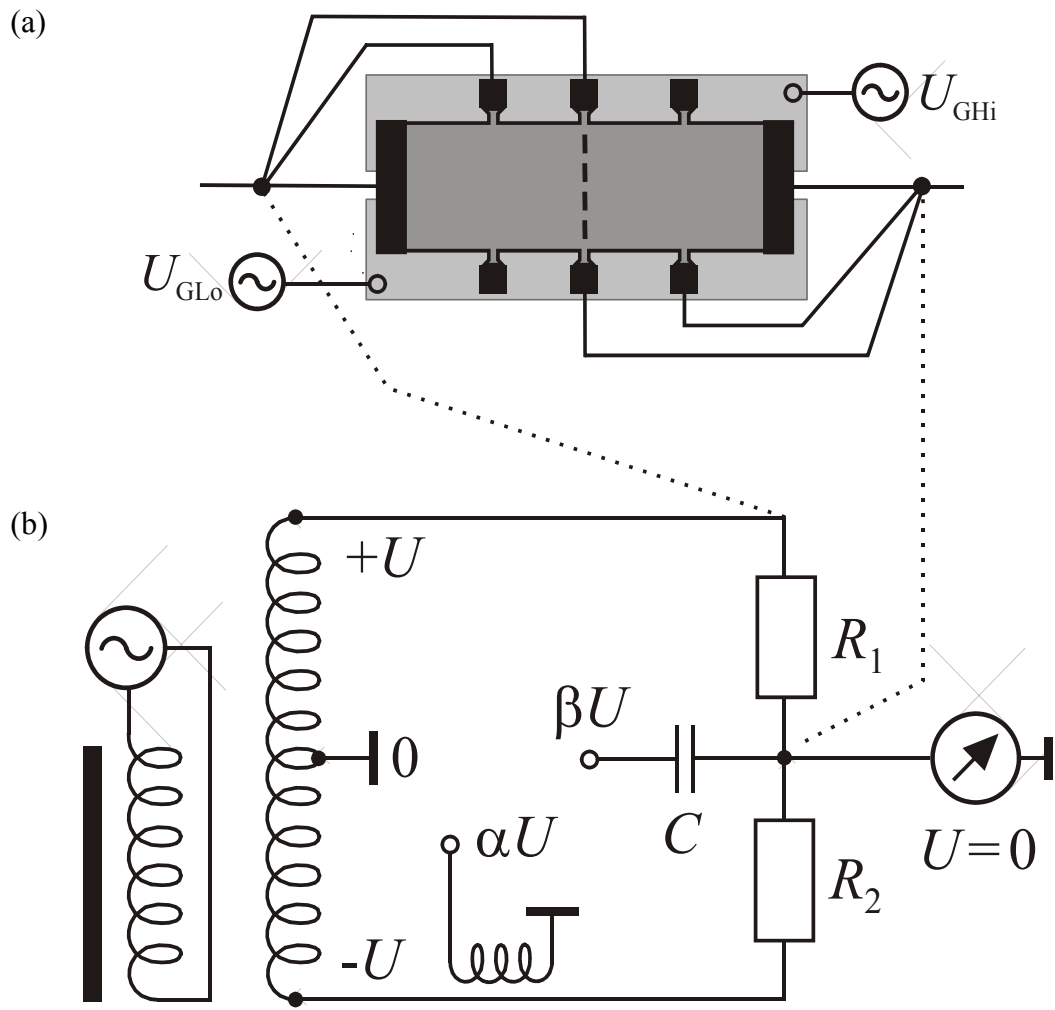

(b)

**Figure 1**

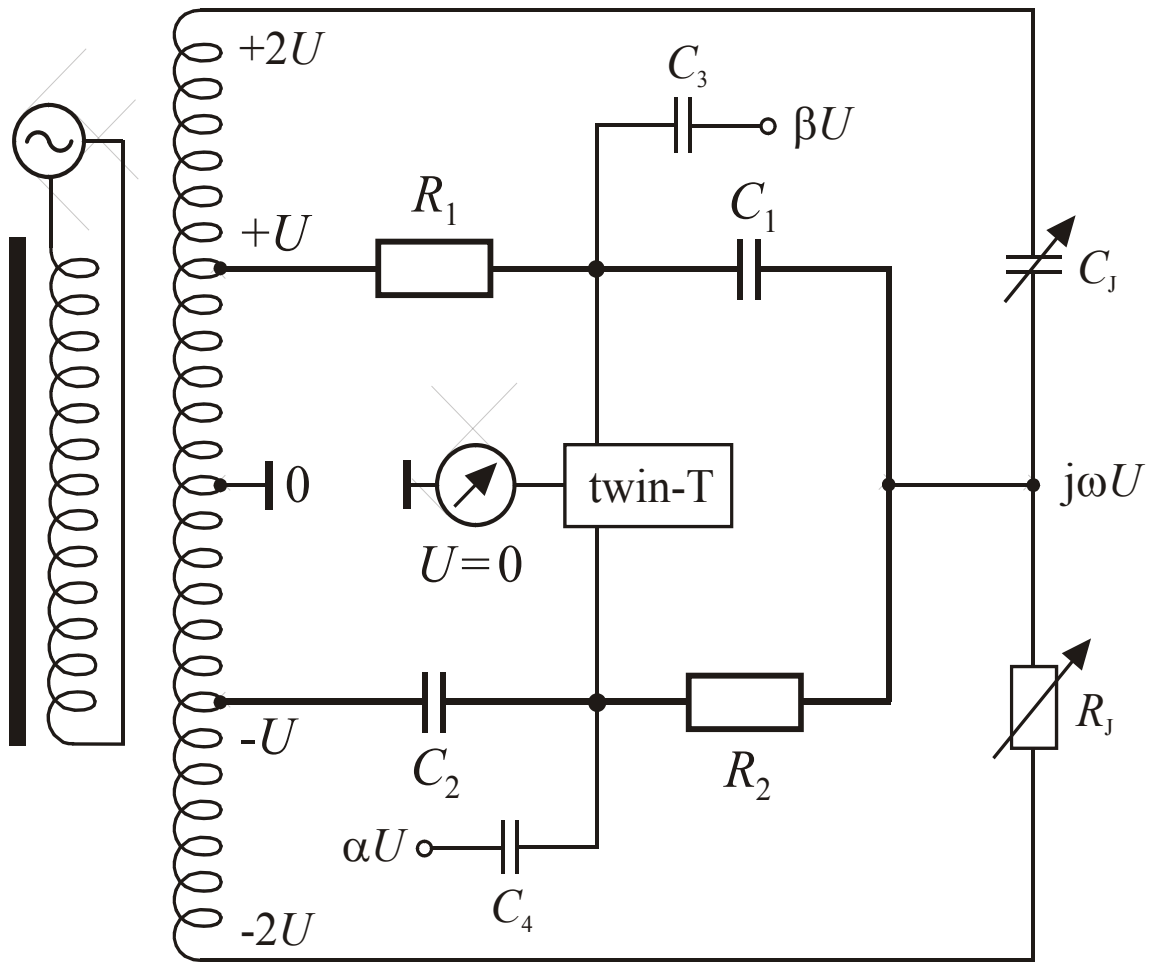

**Figure 2**

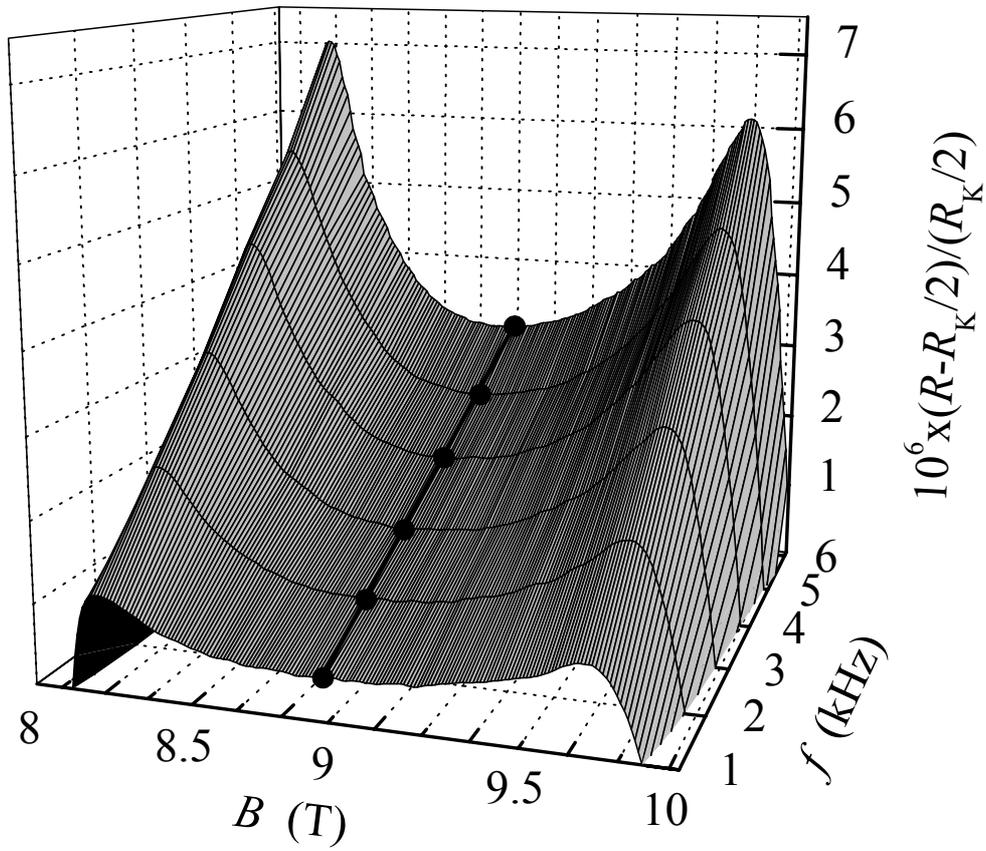

$10^6 \times (R - R_K/2)/(R_K/2)$

$f$ (kHz)

$B$ (T)

**Figure 3**

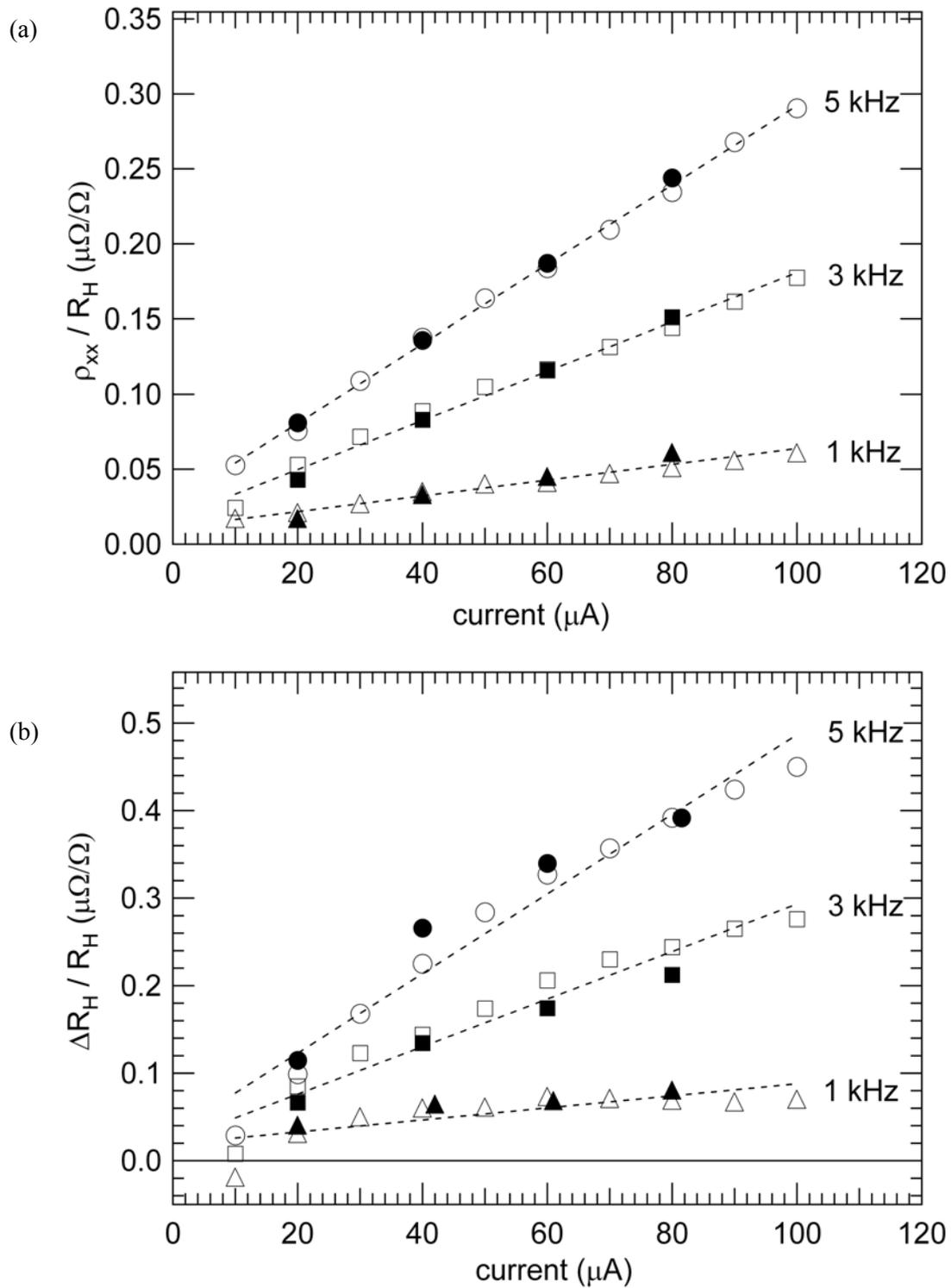

**Figure 4**

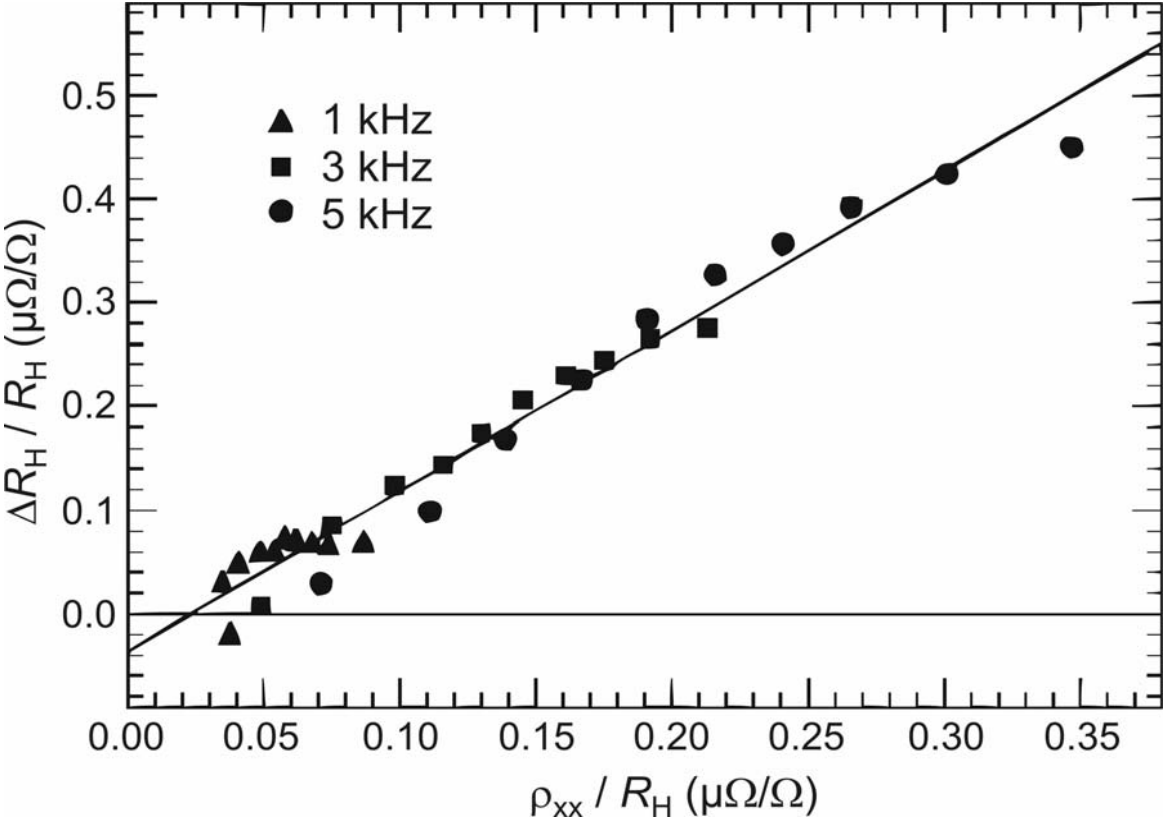

**Figure 5**

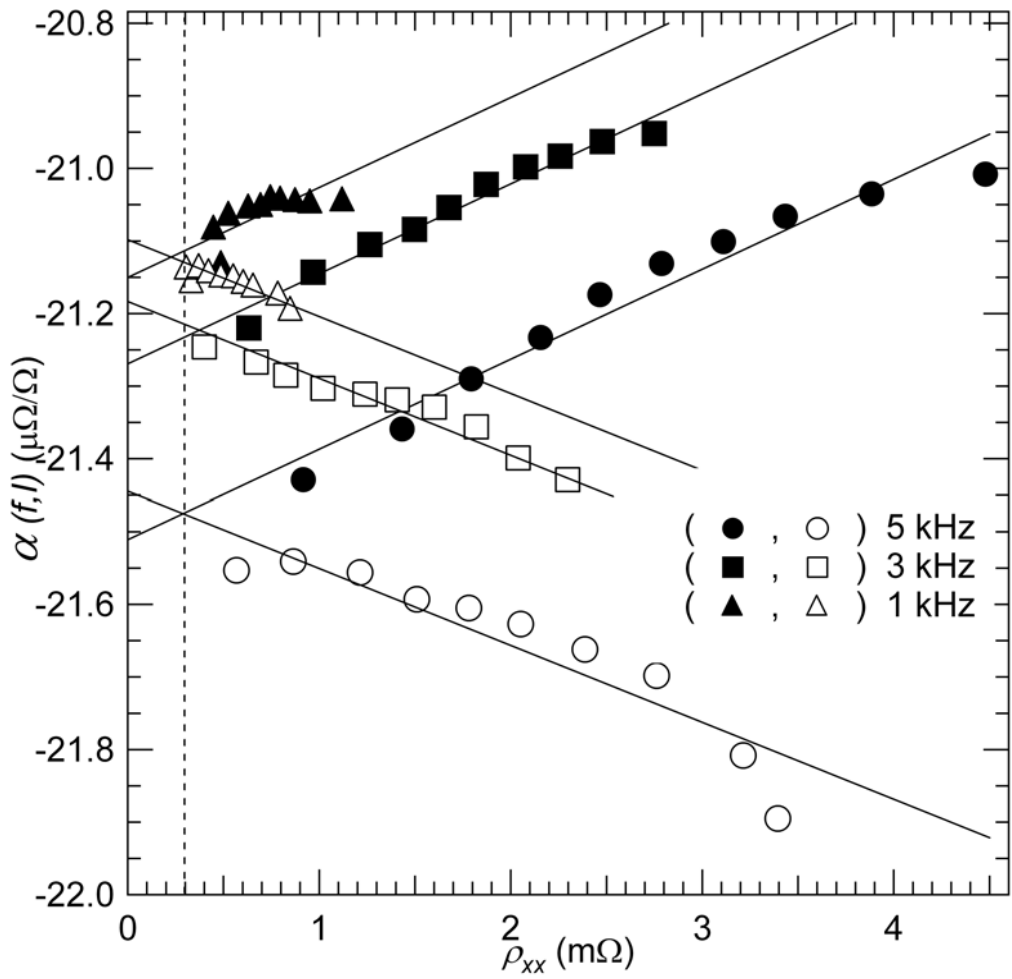

**Figure 6**

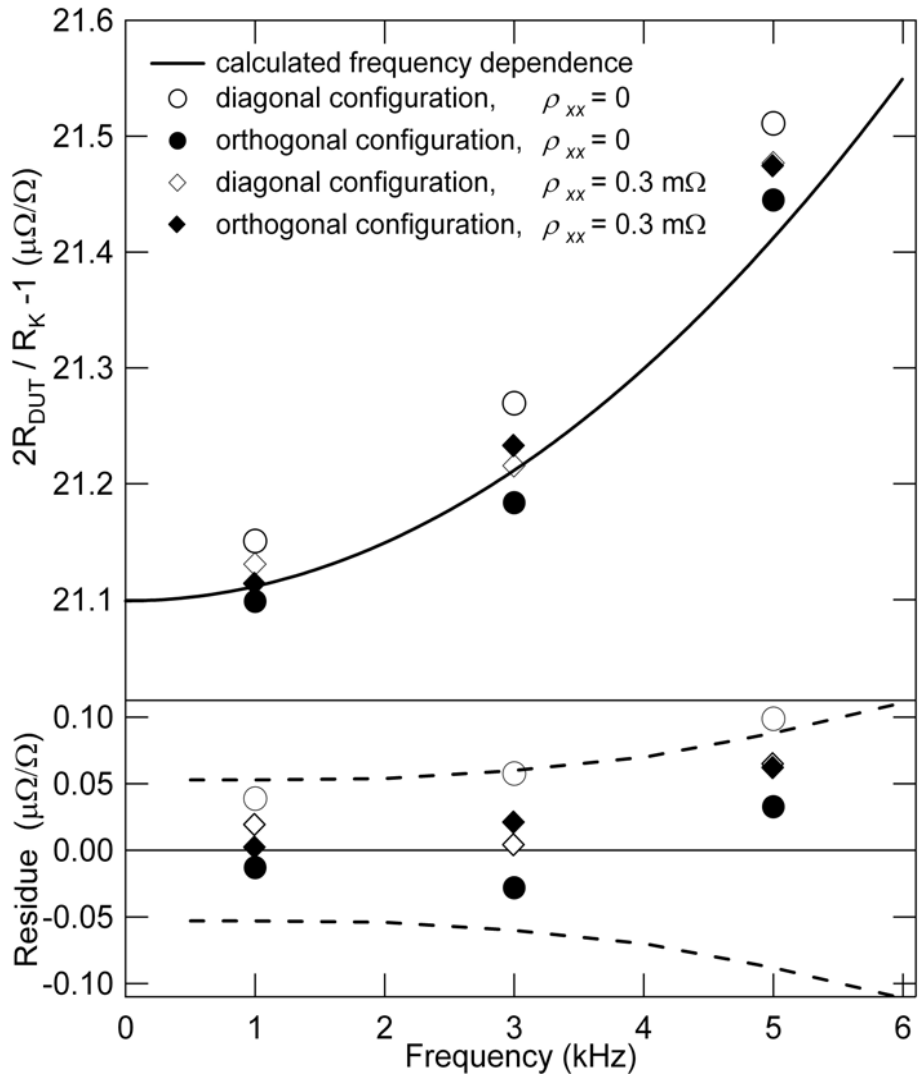

**Figure 7**

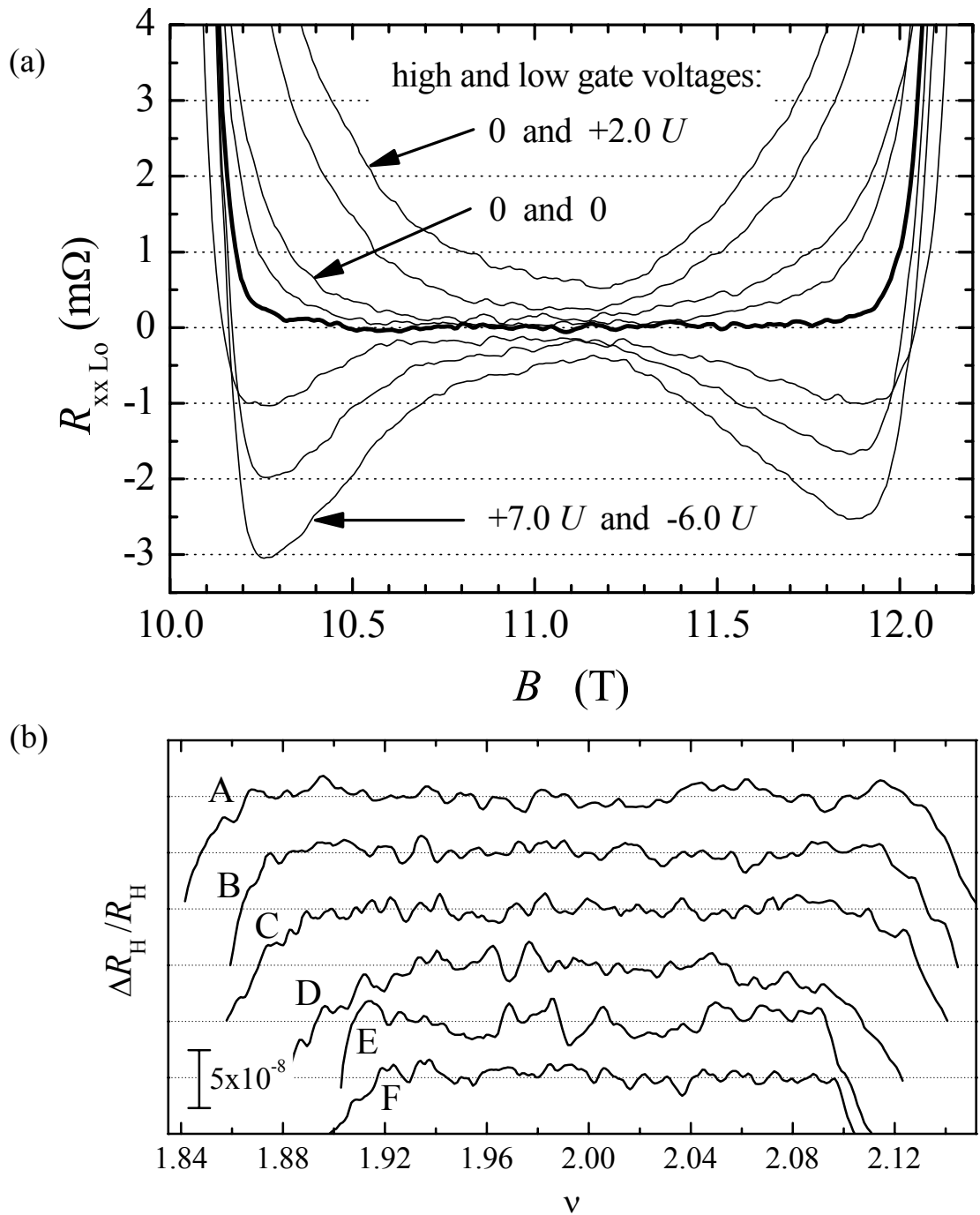

(a)

high and low gate voltages:

0  and  +2.0 $U$

0  and  0

+7.0 $U$  and  -6.0 $U$

(b)

A

B

C

D

E

F

$5 \times 10^{-8}$

**Figure 8**